\documentclass[iop]{emulateapj}
\usepackage{apjfonts}
\usepackage{amssymb}
\usepackage{natbib}
\usepackage{appendix}

\begin{document}

\title{Tracing the reionization epoch with ALMA: [CII] emission in $z \sim$7 galaxies }

\author{L. Pentericci\altaffilmark{1}, S.Carniani\altaffilmark{2,3}, M. Castellano\altaffilmark{1}, A. Fontana\altaffilmark{1}, R. Maiolino\altaffilmark{2,3}, L. Guaita\altaffilmark{1}, E. Vanzella\altaffilmark{4}, A. Grazian\altaffilmark{1}, P. Santini\altaffilmark{1},  H. Yan\altaffilmark{5}, S. Cristiani\altaffilmark{6}, C. Conselice\altaffilmark{7}, M. Giavalisco\altaffilmark{8}, N. Hathi\altaffilmark{9} and A.  Koekemoer\altaffilmark{10}}

\altaffiltext{1}{INAF, Osservatorio Astronomico di Roma, via Frascati 33, 00040 Monteporzio, Italy}
\altaffiltext{2}{Kavli Institute for Cosmology, University of Cambridge, Madingley Road, Cambridge CB3 0HA, UK }
\altaffiltext{3}{Cavendish Laboratory, University of Cambridge, 19 J. J. Thomson Ave., Cambridge CB3 0HE, UK}
\altaffiltext{4}{INAF, Osservatorio Astronomico di Bologna, via Ranzani 1, I-40127 Bologna, Italy }
\altaffiltext{5}{Department of Physics and Astronomy, University of Missouri, Columbia, MO, USA}
\altaffiltext{6}{INAF, Osservatorio Astronomico di Trieste, Via G. B. Tiepolo, 11, I-34143 Trieste, Italy}
\altaffiltext{7}{University of Nottingham, School of Physics and Astronomy, Nottingham NG7 2RD, UK }
\altaffiltext{8}{Astronomy Department, University of Massachusetts, Amherst, MA 01003, USA 0000-0002-7831-8751}
\altaffiltext{9}{Aix Marseille Université, CNRS, LAM (Laboratoire d'Astrophysique de Marseille}
\altaffiltext{10}{Space Telescope Science Institute, 3700 San Martin Drive, Baltimore, MD 21208, USA}

\begin{abstract}
We present new results on [CII]158$\mu m$  emission from four galaxies in the reionization epoch. 
These galaxies were previously confirmed to be at redshifts between 6.6 and 7.15 from the presence of  the Ly$\alpha$ emission line 
in their spectra. The Ly$\alpha$ emission line is redshifted  by 100-200 km s$^{-1}$ compared to the systemic redshift given by the [CII] line. These velocity offsets are smaller than what is observed in $z \sim3$ Lyman break galaxies with similar UV luminosities and emission line properties. Smaller velocity  shifts reduce the visibility 
of Ly$\alpha$ and hence somewhat alleviate  the need for a very neutral IGM at 
$z \sim$ 7 to 
explain the drop in the fraction of Ly$\alpha$ emitters observed  at this epoch.  The galaxies show [CII] emission with L[CII]=$0.6-1.6 \times10^8 L_\odot$: these  luminosities  place them consistently  below   the  SFR-L[CII] relation  observed for  low redshift star  forming and metal poor galaxies and also  below  z=5.5 Lyman break galaxies with similar star formation rates. We argue that previous undetections of [CII] in $z \sim$7  galaxies with similar or  smaller star formation rates are due to  selection effects: previous targets were mostly strong Ly$\alpha$ emitters and therefore probably metal poor systems, while our galaxies are more representative of the general high redshift star forming population.

\end{abstract}


\section {\bf Introduction }
We have just entered an exciting era when cosmic microwave background observations
can be directly compared to observations of the first galaxies. Reionization is
thought to begin at $z \sim 10-15$ and be completed by $z \sim 6$
\citep{rob15}, but how exactly it proceeded in time is still the subject of  debate. Understanding the nature of the sources responsible for such process  is also an outstanding problem of modern cosmology: both faint galaxies (e.g. Finkelstein et al. 2015)  and  AGNs (\citealt{gia15})  can potentially contribute to reionization and the exact role of the two populations is still unclear.

Deep multi-band imaging surveys identified  a large number of candidate galaxies at $z > 6$ and up to $z \sim 10$  primarily  using the Lyman Break technique but their spectroscopic confirmation has been  difficult. At present very few galaxies at  $z>7$ are confirmed (e.g. Finkelstein et al. 2013, Oesch et al. 2015, Zitrin et al. 2015).
Indeed the extreme  difficulty in securing the redshifts of z-dropouts  
is the first (and perhaps most solid) evidence that the reionization of the Universe was not yet complete at $z\simeq 7$. This marked decrease  of Ly$\alpha$ emission at  $z\sim 7$ compared  to $z \sim 6$  is best explained by an increased opacity of the intergalactic medium (IGM)  with  neutral hydrogen fraction change between the two epochs of  $\Delta \chi \sim 0.5$   (Pentericci et al. 2014, Schenker et al. 2014). 
The physical properties of these galaxies are also uncertain. Star formation rates (SFR) based on UV-luminosity give values of $\sim10-15~M_{\odot}$ yr$^{-1}$. However dust is surely present even if in small amounts, at such early epochs (Watson et al. 2015) and can strongly suppress the
UV-continuum. Accounting for this may easily
raise the inferred SFR to several tens $M_\odot yr^{-1}$.  

ALMA can play a key role in settling  the above issues and assess
the nature of high redshift galaxies.  The [CII]157.74$\mu$m line, the  strongest FIR emission lines in star forming  galaxies (accounting for ~0.1--1\% of their bolometric luminosity), is accessible by ALMA even at the highest redshifts probed to date. [CII] can provide SFR estimates that are not  biased by dust extinction,  although they might depend on the  metallicity.   In addition this line allows us to accurately measure the systemic redshift of the galaxies.
The recent [CII] line detections of Lyman Break galaxies (LBGs) and Ly$\alpha$ emitters (LAEs) at  $z \sim$5.5-6 (Capak et al. 2015,  Willot et al. 2015) show that the line  properties at such high redshifts  are  similar to those at  lower redshift  and  that they follow  a comparable  SFR-L[CII] relation ( de Looze et al. 2014). The  observed galaxies  represent the bright end of the UV luminosity function. On the fainter end,   
 Knudsen et al. (2016)  detected [CII] in the strongly lensed  $z=6$ galaxy A383-5.1.
At variance with the above picture,  the few  observations attempted  on  
$z \sim7$ galaxies with ALMA provided contradictory results (e.g. Ota  et al. 2012, Schaerer et al. 2014, Maiolino et al. 2015, Watson et al. 2015):  several  non-detections of [CII]  suggested that $z \sim 7$ galaxies have  [CII] emission significantly  below the expectations from lower redshift  relations, although the small number  of sources observed and the shallow  limits reached for some  of them  give an uncertain scenario. We previously reported of [CII] observations on three $z \sim 7$ galaxies  (Maiolino et al. 2015): for two of them we  set quite stringent constraints on 
the non-detection of the [CII] line, whereas the third one showed a detection from a region that is not  centered on the  galaxy.
None of the $z \sim 7$  galaxies are detected in the far-IR continuum, which suggests a low dust mass. The only exception is A1689-zD1 at $z \sim 7.5$  with a clear detection of thermal dust emission (Watson et al. 2015).
The above results suggest a change in galaxies' physical properties between $z \sim$6 and $z \sim 7$. \\ 
 In this work we present new observations of [CII] in four  LBGs at $z\sim 7$.
Throughout the paper, we adopt a cosmology with $\Omega_{\Lambda} = 0.7$, $\Omega_M = 0.3$ and 
$H_0$=70 km/s/Mpc. Magnitudes  are  in the AB system. 

\section{Sample selection and observations}
We have recently completed CANDELSz7,  an ESO spectroscopic large program using  FORS2 at the VLT. Our goal is to systematically 
 study the Ly$\alpha$ emission in galaxies from z=5.5 to z=7.2 selected in the CANDELS fields (Grogin et al. 2011, Koekemoer et al. 2011).  With the addition of previous  data we have assembled a sample of $>$120 LBGs  at $z\sim7$  with  homogeneous selection  and deep spectroscopic observations (Pentericci et al. in preparation). In  about 20\% we detect the Ly$\alpha$ line, with EW in some cases as low as 5 \AA.  From this sample we  have selected the seven brightest galaxies  with: (1)  precise  redshifts between 6.6 and 7.2 
from Ly$\alpha$ emission; (2)  $SFR > 15-30 M_{\odot} yr^{-1}$  based on UV emission, assuming the Kennicutt (1998) calibration with no dust correction. 
While these LBGs all have Ly$\alpha$ emission, the majority of them would not be selected by the usual LAE criteria because the lines only have modest EW. 

During Cycle-3 we  obtained observations for four of the seven galaxies approved in program 2015.1.01105.S.  In Table 1 we show their optical properties. NTTDF6345 was part  of one of our early studies (Pentericci et al. 2011),  while the other galaxies were confirmed by CANDELSz7. COSMOS13679 has been independently confirmed by Stark et al. (2016).
In Table 1 the coordinates  reported are the HST H-band centroids in all cases except for NTTDF6345, which was detected  with HAWK-I in Y-band (Castellano et al. 2010). For NTTDF6345 we had previously obtained lower S/N data in Cycle-2 (prog. 2013.1.01031.S ).

\begin{table*}[tbh]
\begin{center}
\caption[]{\em{Galaxies  optical and spectroscopic properties}}
\begin{tabular}{lccccccccc}
\hline \noalign {\smallskip}
ID & RA & Dec & redshift & SFR               & Ly$\alpha$ EW   & $M_{UV}$ &  $M_{SED}$  & ref  \\
   &    &     &          & $M_\odot yr^{-1}$ & \AA             &     & $10^9 M_\odot$ & \\ 
\hline \noalign {\smallskip}
COSMOS13679    & 150.0990366   & 2.3436268        & 7.1453  & 23.9   &  15  & -21.46  & 3.0 & 1,2 \\
 NTTDF6345     &  181.4039006  & -7.7561900      & 6.701   & 25.0   &  15  & -21.57  & -- &   3 \\ 
 UDS16291      &  34.3561430    & -5.1856263       & 6.6381  & 15.8     &   6  & -20.97  & 0.6  & 1 \\
COSMOS24108    & 150.1972224   &  2.4786508       & 6.6294  & 29     &  27  & -21.67  & 3.9 & 1 \\
%\hline \noalign {\smallskip}
\end{tabular}
\tablecomments{Redshift reference: 1 Pentericci et al. in preparation; 2 Stark et al. (2016); 3 Pentericci et al. (2011)}
\end{center}
\end{table*}

\begin{figure}
\figurenum{1}
\epsscale{0.95}
\plotone{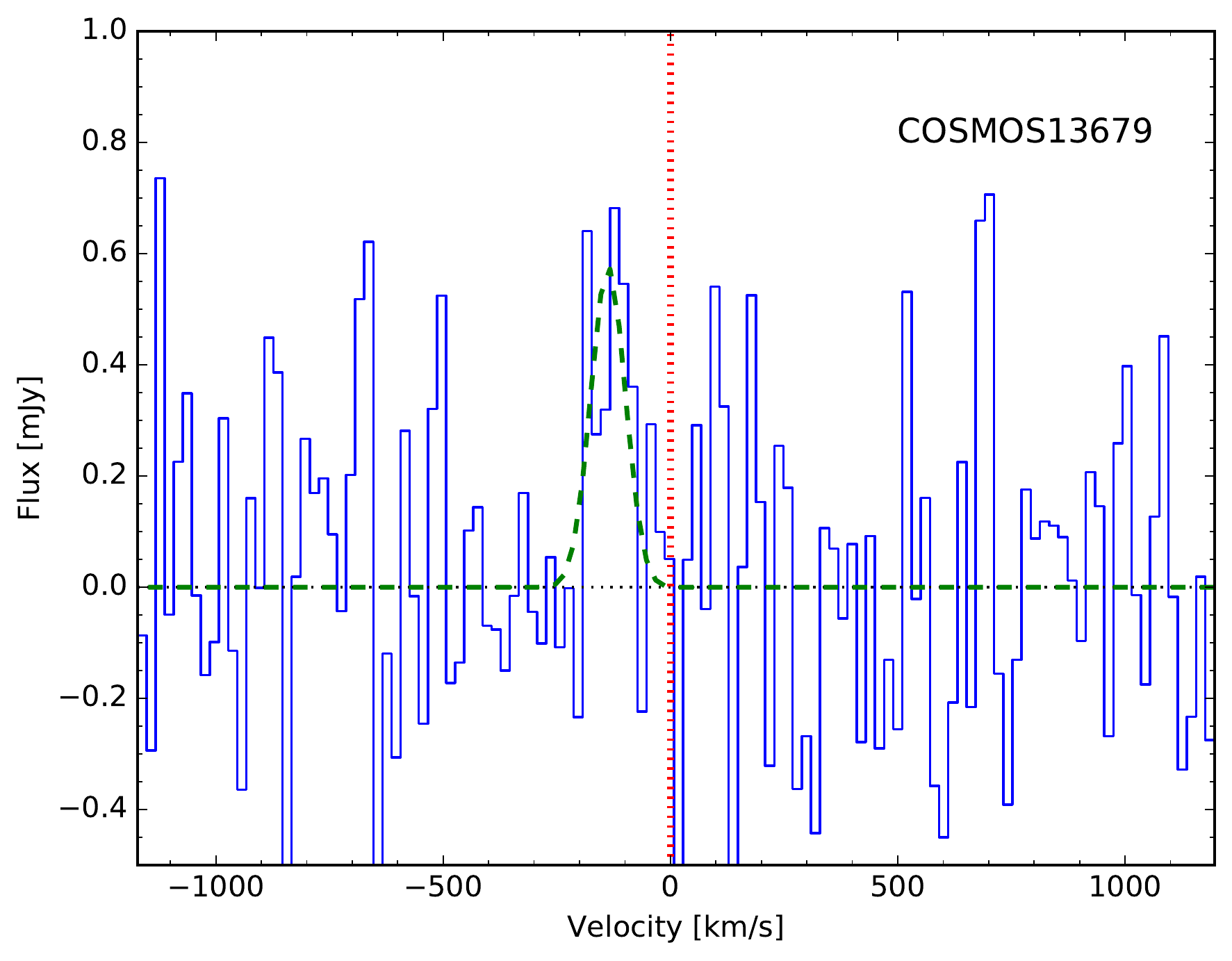}
\plotone{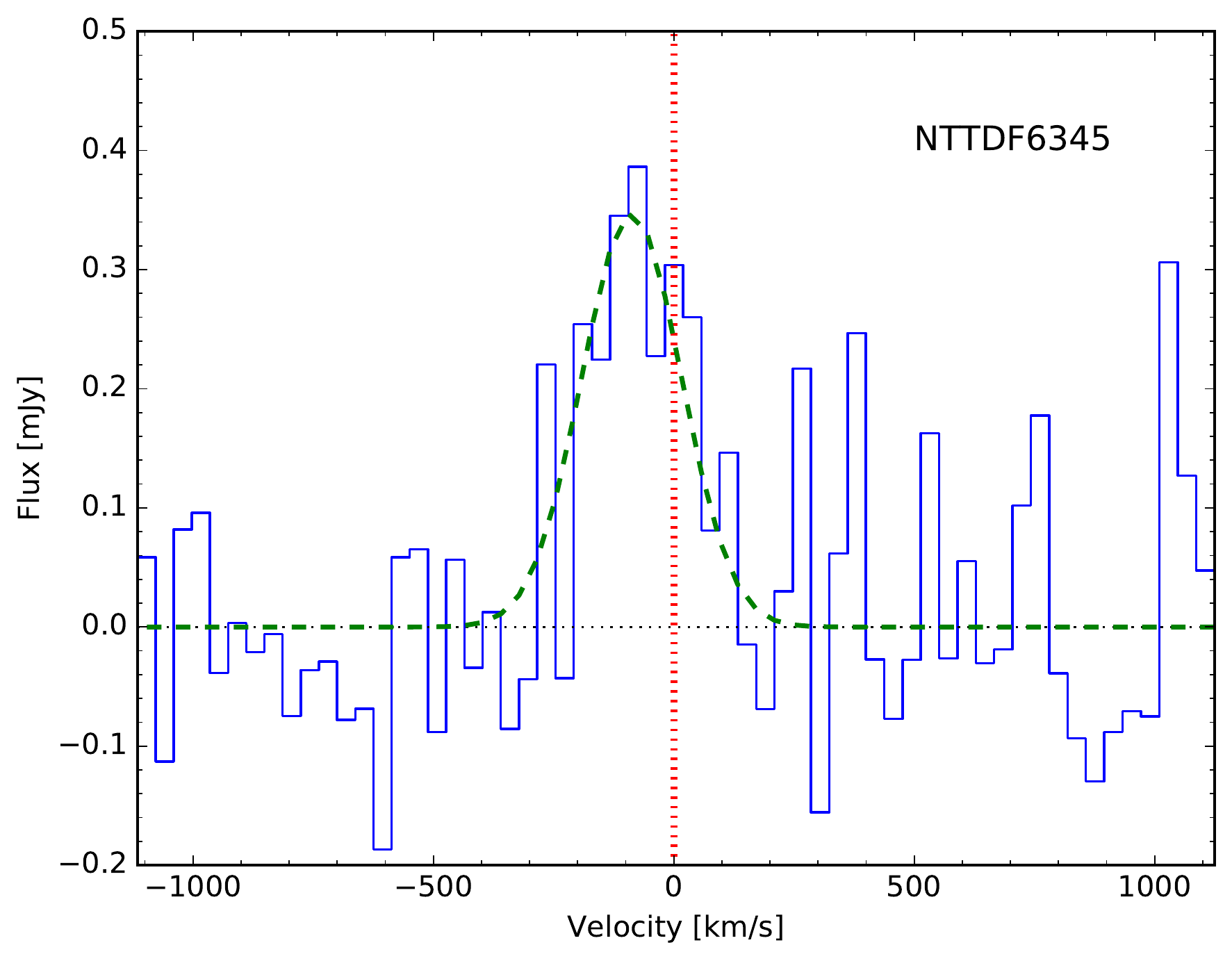}
\plotone{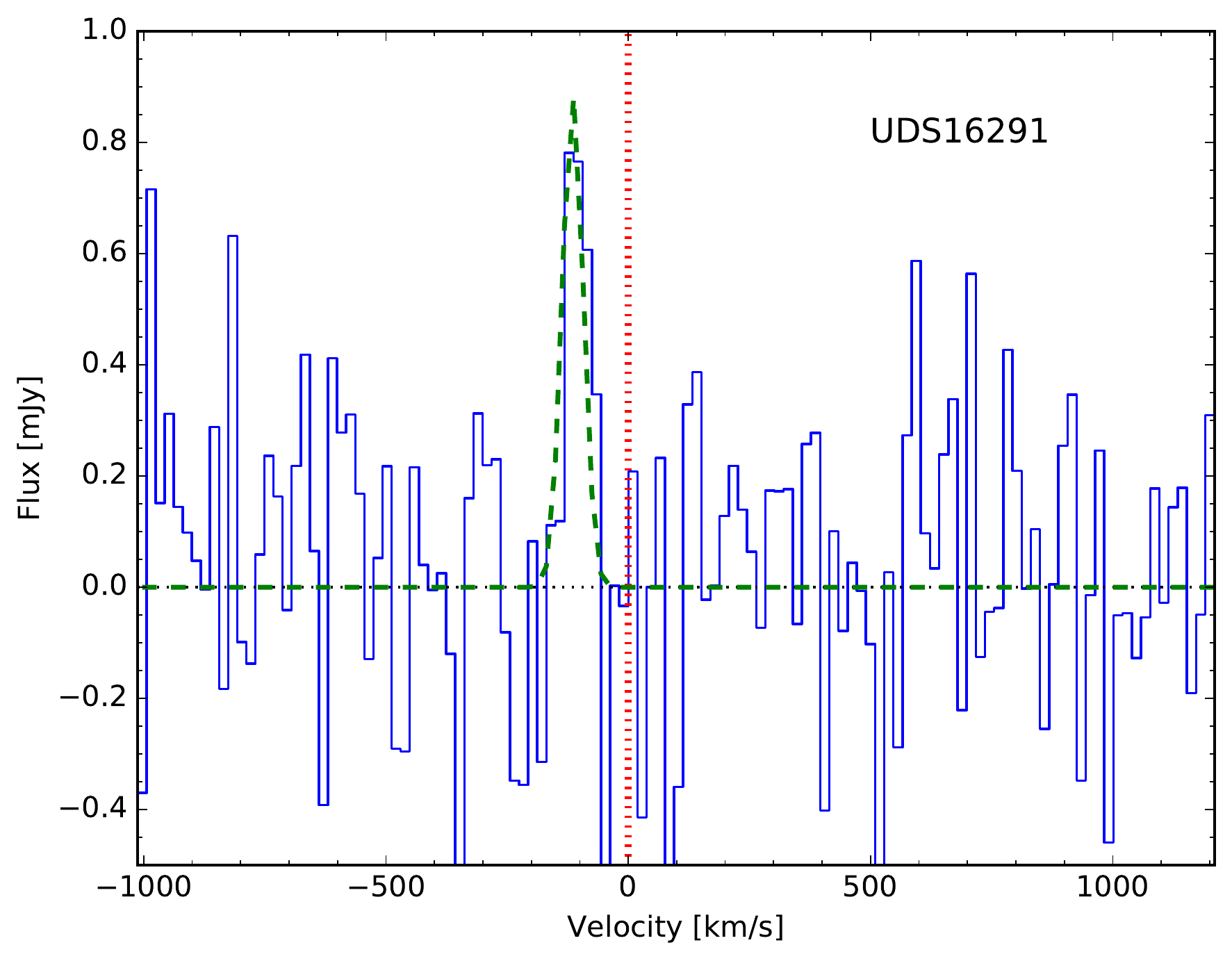}
\plotone{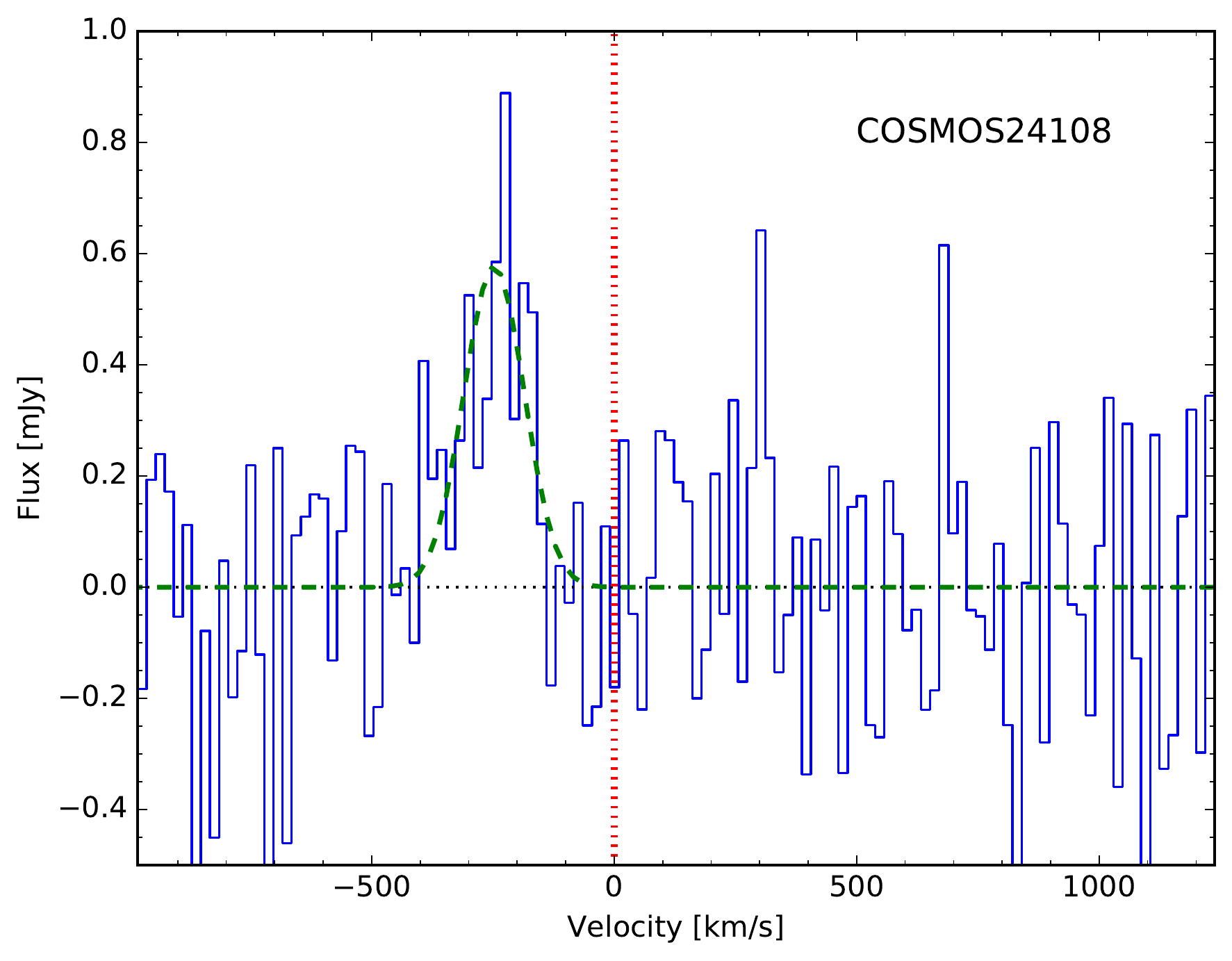}
\caption{S.\label{fig:f1}  
Spectra of the four galaxies in the region of the [CII] emission line. The velocities are shown with respect to that inferred from the Ly$\alpha$  which is marked with the dashed vertical lines.
  }
\end{figure}

Objects were observed with ALMA   in band 6   which has baselines between 14.7 and 376.9 meters, and provides a minimal resolution of 0.9$"$.
The center of the band was set  at the redshift corresponding to the peak of  the Ly$\alpha$  emission.
Sensitivity was set  to reach $\sim$log(L[CII])=7.5 (with L[CII] in $L_\sun$).
Observations were performed in Frequency Division Mode. Out of the four spectral windows, 
SPW1 was  centred on the expected frequency of the [CII] line in the Upper Side Band. This spectral band was set to a spectral resolution of 10 km s$^{-1}$. SPW0 was located on the continuum next to SPW1 (on the higher frequency side), while SPW2 and SPW3 were located in the Lower Side Band to sample the continuum.
\\
The ALMA observations were carried out from April 2015 to March 2016. The number of antennas ranged from 36 to 46.   The precipitable water vapour during the observations ranged between 0.82  and 3.0 mm.
The phases were centred at the  positions reported in Table 1.
The data were reduced with the Common Astronomy Software Application (CASA) and the final images  were produced using the CLEAN task. 
The continuum image of each target was extracted using all the line-free channels of the four spectral windows, while spectral cubes were generated from  the SPW1 dataset.

\section{Results}  
\subsection{[CII] line detections} 
In Table 2 we present our  results:  a [CII] emission line is detected for the three sources observed only in Cycle-3, with a  S/N=4.5. 
For NTTDF6345 an  emission line is observed separately in the Cycle-2 and Cycle-3 data-sets with a S/N of 4.1 and 5.6 respectively, and a S/N>6 is obtained 
in the combination.
Although the S/N is modest,  the spatial coincidence  (or very  close spatial association)  between the [CII]  and the near-IR  counterparts, and the consistent  small shift  with respect to  the redshift determined from the Ly$\alpha$  emission  all argue for the reality of the detections. 
In addition  in three out of four cases the detections are spatially resolved.
\begin{table*}[tbh] 7,635222518 7,623240776
\begin{center}
\caption[]{\em{Galaxies  ALMA properties}}
\begin{tabular}{ccccccccccc}
\hline \noalign {\smallskip}
ID & $\lambda_0$[CII])&  [CII] flux   &  FWHM [CII]    & S/N  & $rms_{cont}$ &   $t_{int}$ &$\Delta pos$ &$\Delta v$ & $M_{dyn}*sin(i)^2$ & $SFR_{dust}$\\
     &          mm     & Jy km/s  & km/s            &    &    $\mu Jy beam^{-1}$        & s & arcsec & km $s^{-1}$ & $10^9 M_\odot$ & $M_\odot yr^{-1}$ \\
 \hline \noalign {\smallskip}
COSMOS13679  &  1.28426$\pm$0.00006   & 5.9$\times 10^{-2}$ &   90$\pm$35  & 4.5  & 14  &  2782 & 0.4 & 135  & <2.5 & <6.2 \\
 NTTDF6345   &  1.2143$\pm$0.0002     & 1.6$\times 10^{-1}$ &   250$\pm$70 & 6.1  & 16 &  2087 & 0.0 & 110  & <18  & <5.7 \\
 UDS16291    &  1.20438$\pm$0.00003   & 6.3$\times 10^{-2}$ &   50$\pm$15  & 4.5  & 20  &  2117 & 0.1 & 110  & <0.7 & <6.6 \\
COSMOS24108  &  1.20249$\pm$0.00007   & 9.2$\times 10^{-2}$ &   150$\pm$40 &  4.5 & 18  &  2177 & 0.8 & 240 & <6.5  & <6.2 \\
%\hline \noalign {\smallskip}
\end{tabular}
\end{center}
\end{table*}
In Figure 1 for each source  we present the spectrum of the [CII] region with a rebinning of 40 km s$^{-1}$ for NTTDF6345 and 20 km s$^{-1}$ for the other sources: the vertical dotted line at v=0 km s$^{-1}$ indicates the redshift determined from the Ly$\alpha$ emission. The maps of the line emission are  shown in Figure 2, extracted with a spectral width of 440 km s$^{-1}$ for NTTDF6345, 280 km s$^{-1}$  for
 COSMOS24108, 100 km s$^{-1}$ for UDS16291 and 120 km s$^{-1}$ for COSMOS13679. The black crosses indicates the centroid of the Y-band image for NTTDF6345 and  of the HST H-band images for the other galaxies.

 Thermal far-infrared continuum is not detected in any of the galaxies:  the limits on the total IR-luminosity convert into limits on the dust obscured $SFR_{dust}$ that are reported in Table 2 (assuming a Kennicutt (1998) relation with a Salpeter initial mass function).
\\
\subsection{Offsets between [CII] and rest-frame UV position} 
In Table 2  we report the offset between the near-IR  coordinates and  the ALMA detections.
In two cases the shifts are consistent with the ALMA astrometric uncertainty (0.1-0.15$''$), while for COSMOS24108 and COSMOS13679  they are larger. 
Dunlop  et al.  (2016)  recently noted that the HST and ALMA astrometry of the HUDF field  presented both  a systematic shift of 0.25$"$  to the south and a random shift of up to 0.5$''$.
Spatial offsets of up to 0.5$''$ are clearly  evident   in most of the LBGs observed by Capak et al. (2015)  and for the $z \sim$6 galaxy WMH5 observed by Willot et al. (2015) ($\sim 0.4''$), while in the other galaxy CLM1, the [CII] emission is co-spatial with the UV continuum.

To further investigate  this  issue we looked at bright serendipitous sources  detected in the continuum band in our fields. We find at least one source per field  and   we measure a shift between 0.1$''$ and 0.6$''$
 in random directions between  the HST (or HAWK-I) counterpart and the ALMA detection. Given   the depth of the CANDELS images ($H\sim 27$ at 5$\sigma$) it is unlikely that the ALMA detections are associated to  other objects  undetected in the H-band.
 Note that for NTTDF6345 where we only have HAWK-I images, the shift is negligible both for  the serendipitous source and the LBG. We conclude that there are still substantial uncertainties in the relative ALMA-HST astrometry.  
In summary, for  UDS16291 and NTTDF6345 the [CII] detections are  centered at the position of the near-IR
 sources, and   for COSMOS13679 the offset (0.4$"$) is within the range of those reported in the literature and measured for the serendipitous sources: for these three galaxies we conclude that the [CII] emission comes from the same region as the bulk of the SF  observed in the near-IR images. 
For  COSMOS24108  the  offset is slightly larger than the range reported in the literature, and while the [CII] emission is almost certainly associated to the source, it  could actually  come from an external region of the galaxy, not coincident with the bulk of the star formation. A similar case was  already observed in 
BDF3299 at z=7.109 (Maiolino et al. 2015): for this galaxy we concluded  that the  [CII] emission arises from an  external accreting/satellite clump of neutral gas,  in agreement with recent models of galaxy formation (Vallini et al. 2015).  While the identification with another transition from a foreground galaxy, which by chance happens to be at 0.8$''$ from COSMOS24108, is  unlikely, we cannot completely discard the possibility of a spurious detection given the low S/N.

\begin{figure*}
\figurenum{2}
\epsscale{1.15}
\plottwo{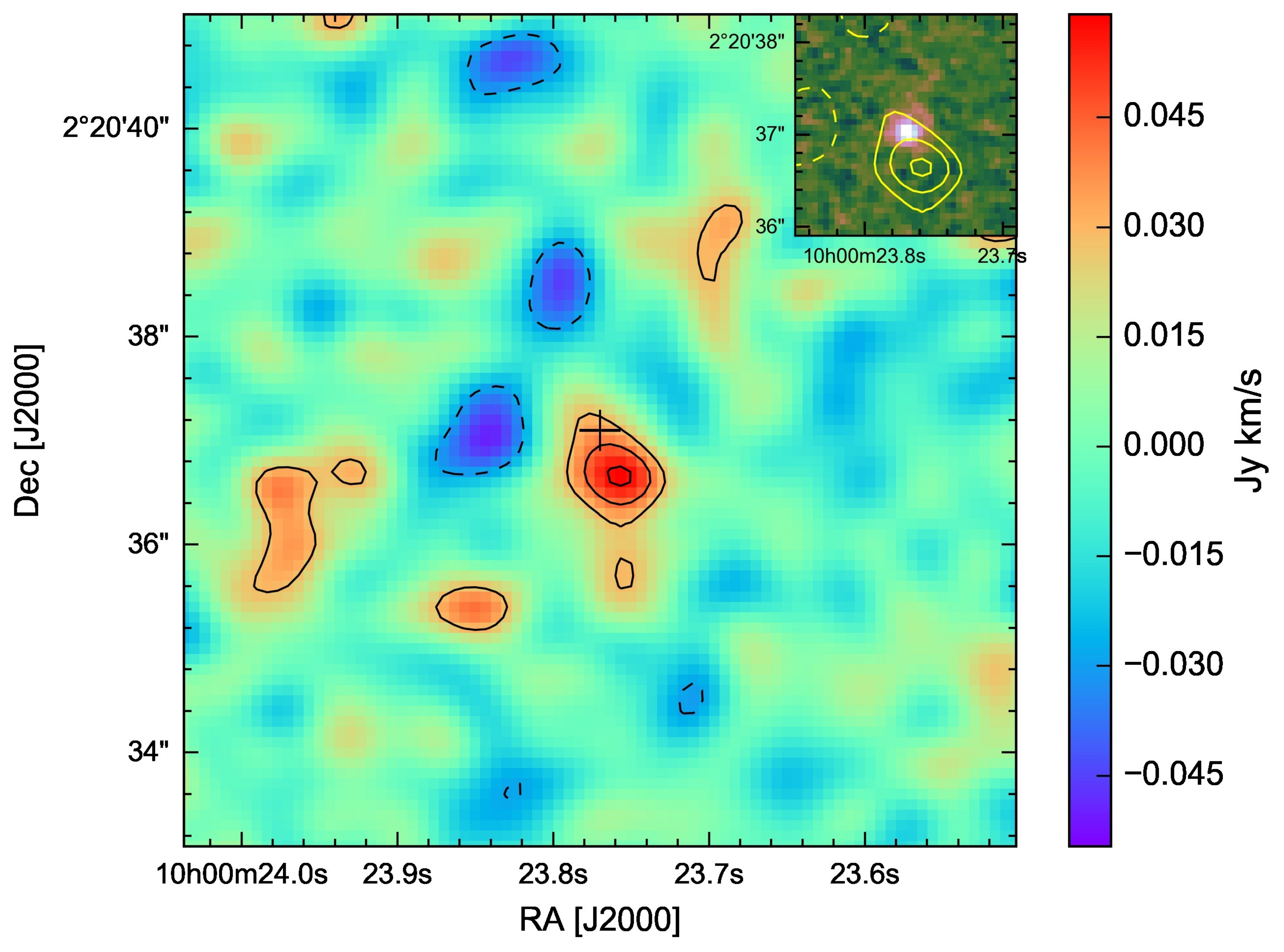}{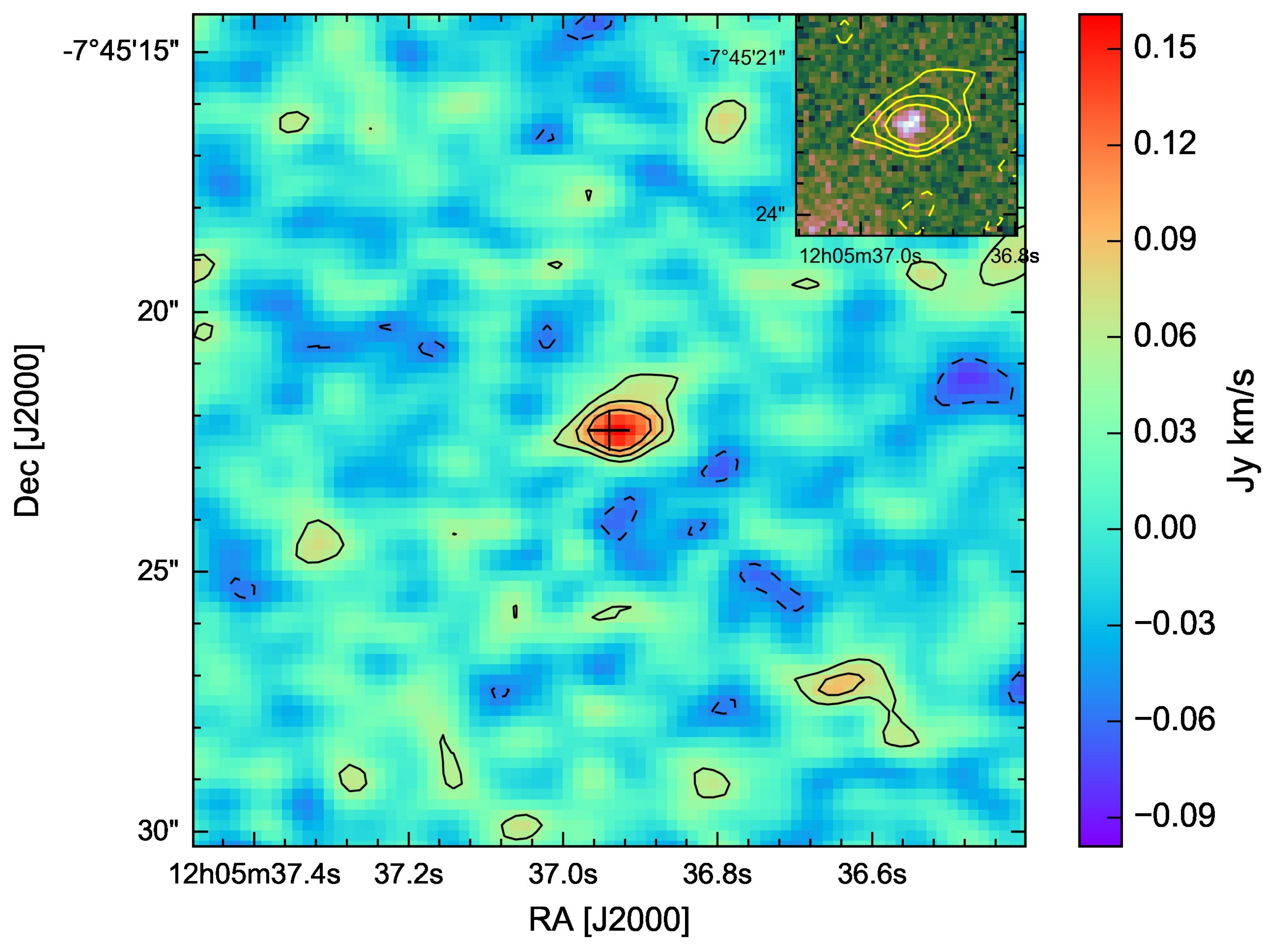}
\plottwo{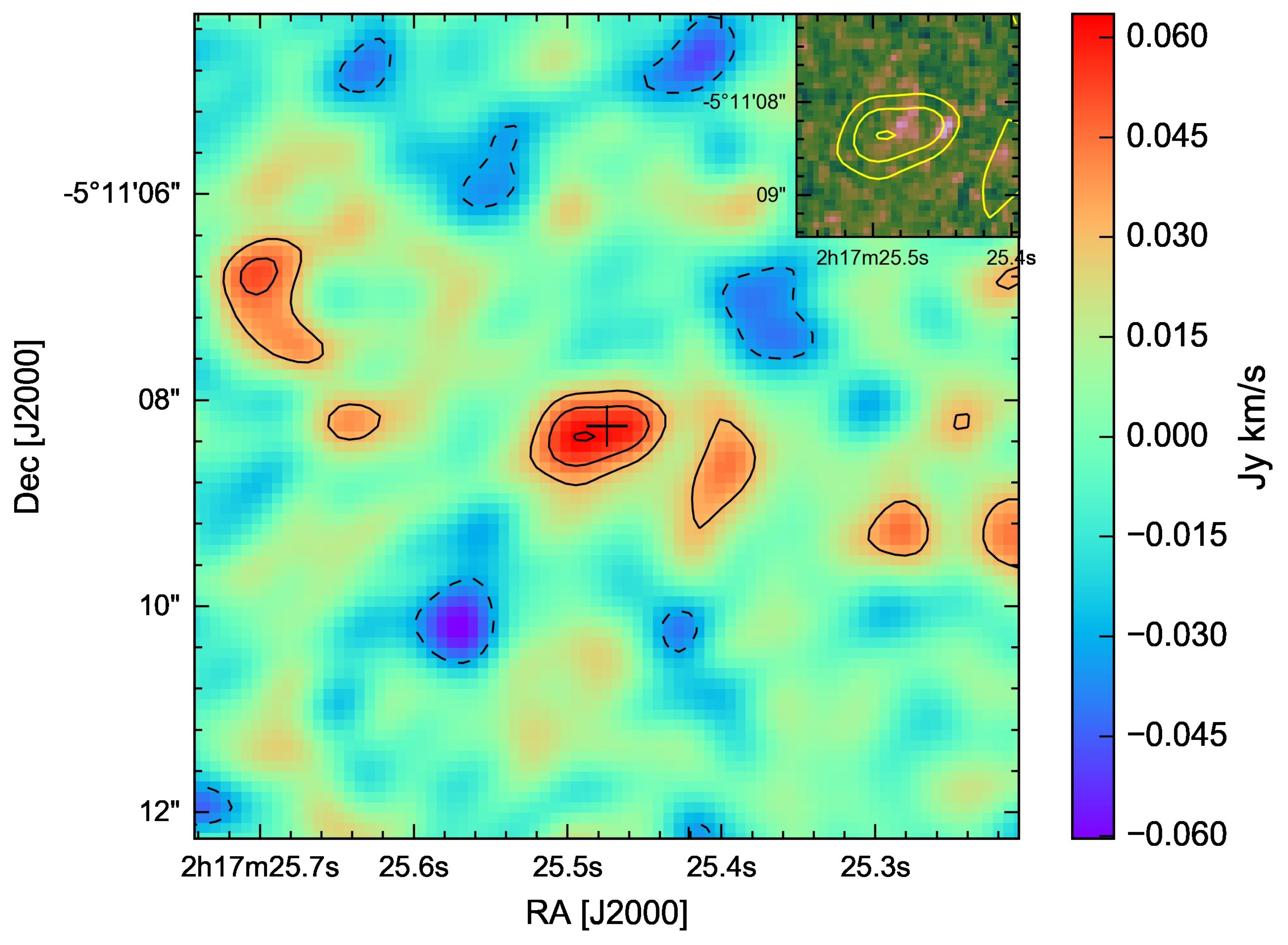}{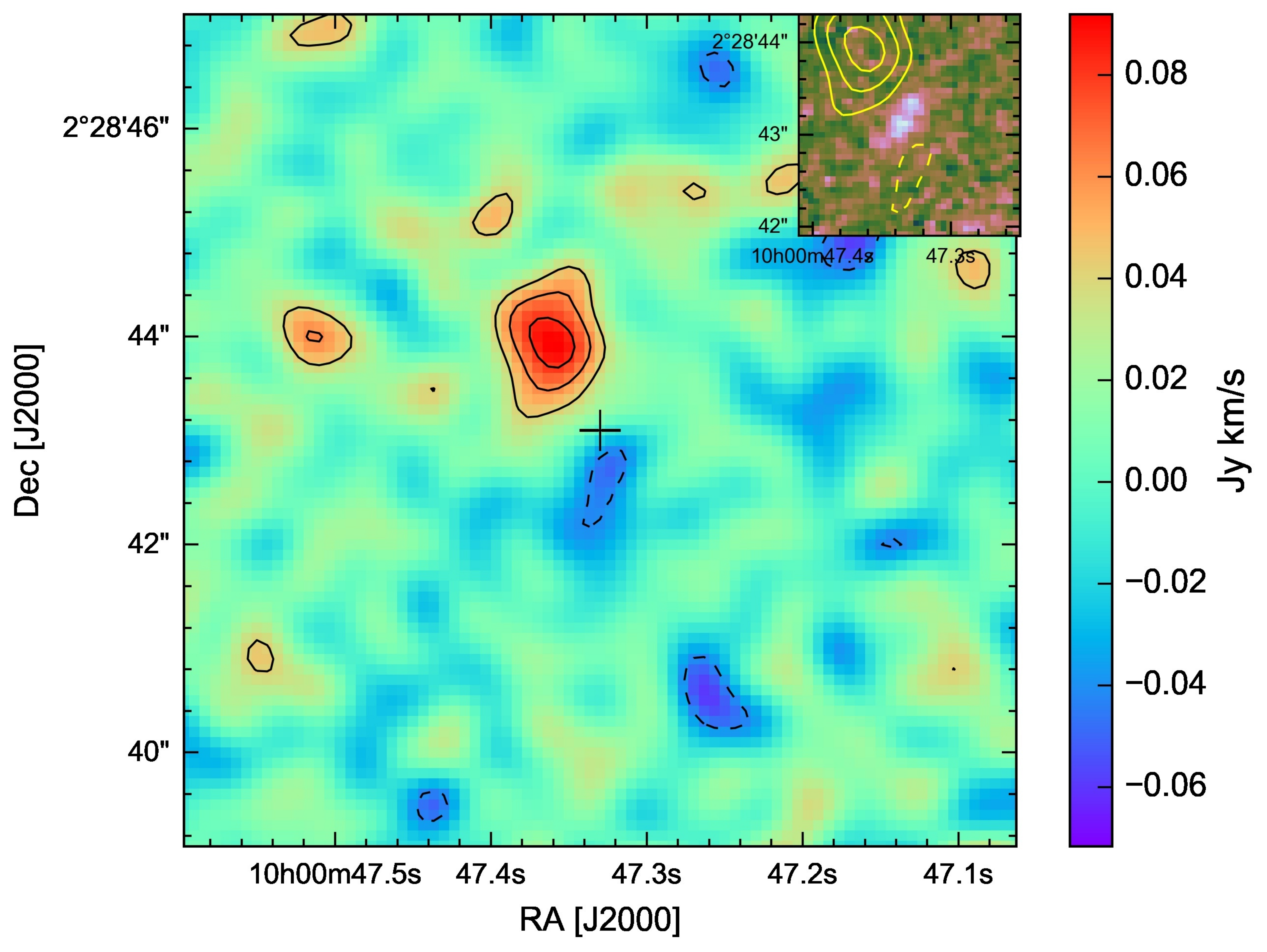}
\caption{\label{fig:f2}  Maps of the [CII] emission for COSMOS13679 (upper-left), NTTDF6345 (upper-right), UDS16291(lower-left) and COSMOS24108 (lower-right). Contours are -2,2,3,4$\sigma$ level. 
The crosses indicate the  position of the near-IR detections. In the insets we show the near-IR images with ALMA contours overlayed. }
\end{figure*}

\section{Discussion}
\subsection{[CII] - Ly$\alpha$ velocity shifts} 
The [CII] line  traces the systemic redshift of the sources,  unlike  the Ly$\alpha$ line which is typically red-shifted by  up to several hundreds  km $s^{-1}$ (Erb et al. 2014, Trainor  et al. 2016) consistent with the presence of outflowing gas, although the final observed Ly$\alpha$ profiles depend on many factors such as geometry, gas covering fraction, dust and IGM ionization state. 
For our galaxies the velocity shifts are not very large,  of the order of 100-200  km s$^{-1}$, 
smaller than those reported at $z\sim3$ for galaxies with similar UV luminosities. Specifically Erb  et al. (2014)  measure shifts of up to 1000 km s$^{-1}$  and average values of 400 km s$^{-1}$ for LBGs with $M_{UV} < -21$. 
The mean shift  is  also lower than those reported by Willott et al. (2015),  430 and 275 km s$^{-1}$  respectively for their two $z \sim$6 galaxies\footnote{For the Capak et al. (2015) sources this comparison is not possible since  their UV redshifts are not from Ly$\alpha$ emission}.
The small  shifts in our galaxies are particularly significant given  that in general objects with low Ly$\alpha$ emission  have larger velocity  offsets.  A similar tentative evidence for smaller Ly$\alpha$ velocities at $z \sim 7$ compared to $z \sim3$  was recently reported by Stark et al. (2015) using the UV nebular CIII]$\lambda 1909$ doublet to determine the systemic redshift in two distant LBGs.  

The velocity  of Ly$\alpha$ compared to the systemic redshift is very relevant when  
interpreting  the line visibility during the reionization epoch, in the presence of a partly neutral IGM \citep{djk11}. Smaller velocity offsets imply that Ly$\alpha$  is closer 
to resonance and more easily quenched by a neutral IGM.
If at redshift 6 and 7  the offsets are as large as those  found for  lower redshift LBGs, the IGM must  be very neutral to  produce the drop in Ly$\alpha$ fraction that is observed between these two epochs  \citep{pen14}. On the other hand, if the velocity offsets were much smaller, as our observations indicate, the drop  in the Ly$\alpha$ visibility  could  be  produced by an IGM that is still substantially more ionized \citep{mes15}.

\subsection{SFR-L[CII]  relation}
In Figure 3  we show  the SFR-L[CII]  relation for our galaxies and  previously observed   sources. We plot COSMOS24108 with a different symbol
because it is not certain whether its [CII] emission is from the main galaxy or just from a clump in its outskirt, in which  case the point would shift $\sim 1 dex$ to the left. 
 We remark that the SFR for our sources as well as previous $z\sim 7$ ones are  UV-based, with no correction for dust extinction. As stated above the upper limits on the $SFR_{dust}$ are very low, at least  for our galaxies. The SFR for the Capak et al. (2015) sample include both the UV and dust obscured contribution. Finally for the  Willot et al. sample we plot SED-derived SFR. 
Our galaxies are a factor of 2-3  less luminous in L[CII] than $z \sim$5.5 galaxies with similar SFRs. They also fall below the SFR-L[CII] relation of low redshift star forming galaxies  and low metallicity galaxies (black solid and dashed lines \citealt{del14}).
\\
Previous observations of $z \sim 7$ galaxies failed to detect the [CII] emission. Some of the galaxies were fainter  than ours, but few others were in the same range and in these cases the limits reached  were  deep enough to detect [CII]  if the emission was at the same level as in  our sources (Maiolino et al. 2015, Schaerer et al. 2015). 

However we note that previously observed sources were either selected as  Ly$\alpha$ emitters or were LBGs but their spectra showed  a Ly$\alpha$ emission with very high  EW  (typically $>$40 \AA). The only exception z8-GND-5296 at z=7.5 which has modest Ly$\alpha$ emission, but in this case the 
L[CII] limit  is very shallow  \citep{sch15}. 
Our four new sources  have Ly$\alpha$  emission with low EW (Table 1).  
The Ly$\alpha$ emission strength  is known to depend on the presence on dust and possibly  metallicity \citep{rai10}: although the  derivation of metallicity is not easy, several studies indicate  that   LAEs are  more metal poor galaxies compared to the rest of the 
LBG population (e.g. Song et al. 2014).
Metallicity plays an important role in shaping the SFR-L[CII] relation:  in Figure 3 we show different metallicity dependent relations produced by a 
recent study of  Vallini et al. (2015) based on high-resolution, radiative transfer cosmological simulations.  The different lines have Z=0.05,0.1 and 0.2 $Z_\odot$ respectively, and for metallicity $\leq 0.1$ are  consistently below the relation found for local  galaxies.
Therefore the contradictory  results for $z \sim7 $ galaxies might be due to different intrinsic metallicity.

Alternatively galaxies could be caught in different evolutionary stages:  it has been suggested that molecular clouds in the central parts of 
primordial galaxies could be rapidly disrupted by stellar feedback hence suppressing the emission of [CII]. Clumps of neutral gas (or small satellites) in the outer regions of such galaxies, could survive photo-ionization and 
still show [CII] emission,  as a consequence of the diffuse far-UV radiation emitted by the primary galaxy.    
This could be the case of  COSMOS24108, where the [CII] emission is  offset from  the near-IR source, 
and could come  from one such clump. A similar occurrence was also observed in BDF3299 at z=7.109 where the [CII] emission was displaced by the near-IR source by 4 kpc.  We refer to \cite{mai15} for an extensive discussion of this scenario.

\begin{figure}
\figurenum{3}
\epsscale{1.5}
\plotone{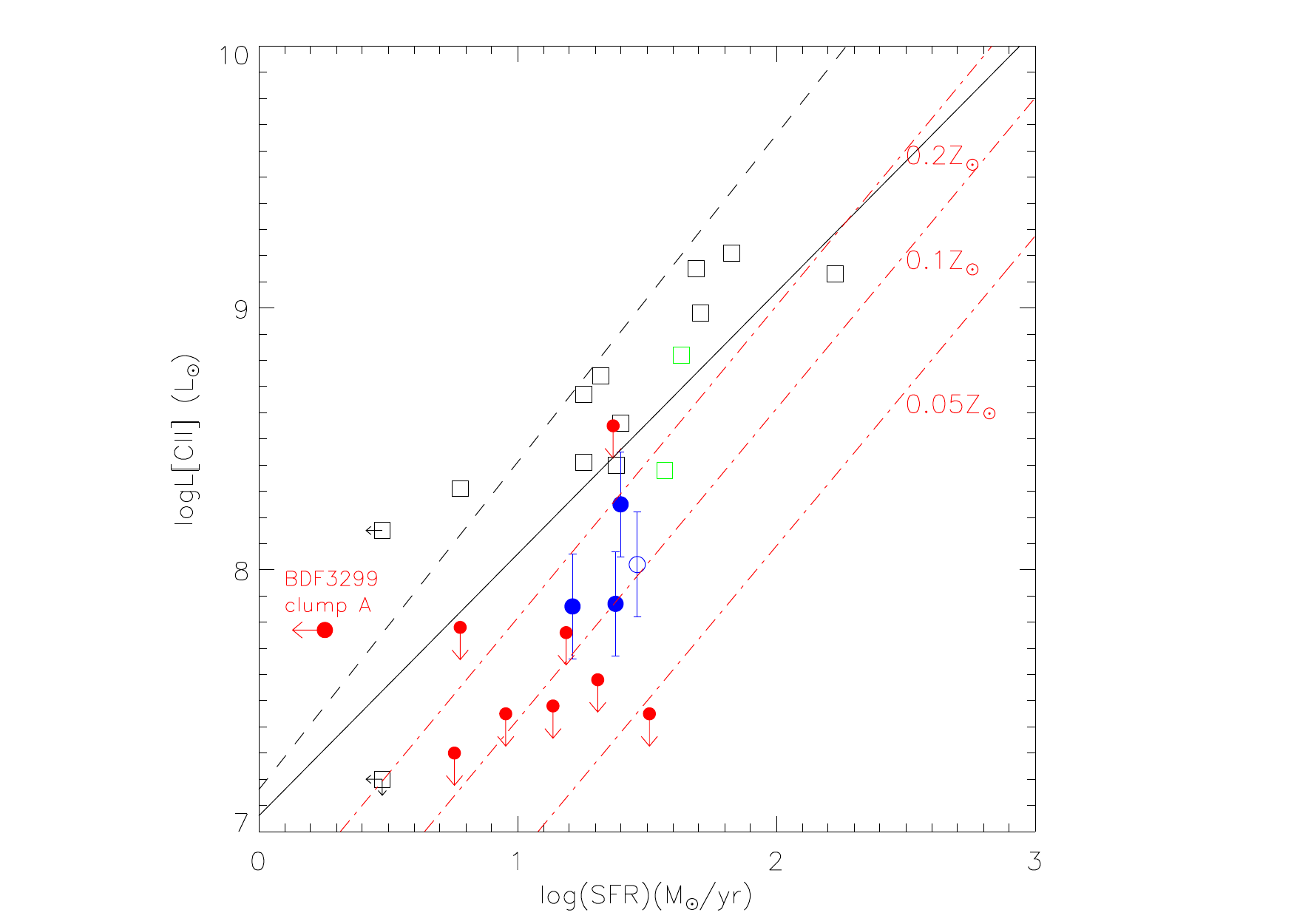}
\caption{S.\label{fig:f3} The blue  circles represent the galaxies  of the present study with the empty circle indicating COSMOS24108 for which the association between the [CII] emission and the optical galaxy is uncertain. Red circles are previous  $z\sim$7 objects (Schaerer et al. 2014;  Maiolino et al. 2015). Empty black  squares are z$\sim$5.5 galaxies from Capak et al. (2015); 
green squares  are z$\sim$6 objects from Willott et al. (2015).
The black lines show  the relations for local star forming galaxies and star-bursts (not-including ULIRGs) (solid) and low redshift metal poor dwarf galaxies (dashed) (De Looze et al. 2014). The  red dot-dashed lines are the resulting relations from simulations \citep{val15} 
for $Z = 0.05 Z_\odot$, $0.1Z_\odot$ and  $0.2 Z_\odot$ respectively.     }
\end{figure}

\subsection{Dynamical masses}
Assuming the sources have ordered motions, we can estimate the dynamical masses based on the [CII] velocity dispersion. 
We follow the method described  in Wang et al. (2013) who approximate  $M_{dyn} = 1.16 \times 10^5 V_{cir}^2 D$, where $V_{cir}$ is the circular velocity in km 
s$^{-1}$, D is the size  in kpc, $V_{cir} = 1.763 \sigma[CII]/sin(i)$, and $i$, the disk inclination angle. Our sources are only marginally resolved so we assume an upper limit on D  from the resolution of 0.9$''$.
We obtain masses between $0.7-18\times 10^9 M_\odot$. We can compare these values to the total 
 stellar masses  determined by a classical  SED fitting of the multi-wavelength photometry for the 3 CANDELS sources. We use the photometry from the CANDELS catalogs,  including deep IRAC data  essential for sampling the rest-frame optical emission at these  redshifts. We include the contribution from   nebular emission lines,  which  can strongly contaminate the IRAC bands.  
The stellar masses are reported in Table 1 and  are very similar to the dynamical masses, especially considering the uncertainty due to the unknown inclination angle $i$. 

\section{Conclusions}
We have presented ALMA  [CII] emission line detections  in four  $z \sim 7$ galaxies. 
Our observations demonstrate that it is possible to detect [CII] during the reionization epoch and that  the line 
luminosity is $\sim 3$ times lower than expected on the basis of the lower redshift relations. 
We find evidence for reduced velocity offsets of the Ly$\alpha$ emission  compared to $z \sim 3$ LBGs: if confirmed this could have important implications for reionization models, since small shifts alleviate the need for a very neutral IGM to reproduce the  observed  decline of Ly$\alpha$ emission at $z \sim7$.
These results  can help us tune future ALMA  observations of high-z LBGs where  no Ly$\alpha$ emission is visible but where precise photometric
 redshifts exist, as in the  CANDELS fields.  Given that in the reionization epoch most  galaxies  do not show Ly$\alpha$ emission, this means that [CII] can be an efficient  alternative to derive the redshift for the majority of the galaxy population.
While  larger samples of targets with different luminosities and Ly$\alpha$ properties are  needed to put our conclusions on firmer grounds,  additional diagnostic can be used to interpret the physical conditions in these objects. 
For example  in case of photo-ionization feedback  [NII] at 205$\mu$m  can be much stronger than [CII]  (Pavesi et al. 2016).  Alternatively the  [OII] line at 88m$\mu$ could be brighter than [CII] in  chemically unevolved systems (Cormier et al. 2015) as  recently observed in a dust-poor galaxy at z=7.2 (Inoue et al. 2016).

\acknowledgments
This paper makes use of the following ALMA data: ADS/JAO.ALMA\#2015.1.01105.S. ALMA
is a partnership of ESO (representing its member states), NSF (USA) and NINS
(Japan), together with NRC (Canada), NSC and ASIAA (Taiwan), and KASI (Republic of
Korea), in cooperation with the Republic of Chile. The Joint ALMA Observatory is
operated by ESO, AUI/NRAO and NAOJ.

This paper is based on data obtained with ESO program 190.A-0685.

SC and RM acknowledge financial support from the Science and Technology Facilities
Council (STFC)
\end{document}